Figures are available from the authors.

\documentstyle[preprint,aps]{revtex}
\begin{document}
\title{ Coulomb plus Nuclear
Scattering in Momentum Space\\ for Coupled Angular-Momentum
States}

\author{Dinghui H. Lu, Tim Mefford, Guilian Song*, and Rubin H. Landau}
\address{Department of Physics, Oregon State University, Corvallis, OR
97331}

\footnotetext {$\mbox{}^{*}$ Current address: Physics Department,
Harbin Normal University, P.R. China.}
\date{\today}
\maketitle

\begin{abstract}
The Vincent--Phatak procedure for solving the momentum-space
Schr\"{o}dinger equation with combined Coulomb-plus-short-range
potentials is extended to angular momentum states coupled by an
optical potential---as occurs in spin $1/2 \times 1/2$ scattering.  A
generalization of the Blatt--Biedenharn phase shift parameterization
is derived and applied to $500$ MeV polarized-proton scattering from
$\mbox{}^{3}$He and $\mbox{}^{13}$C.  The requisite high-precision
partial-wave expansions and integrations are described.
\end{abstract}

\section{Introduction}

The theory and equations of quantum mechanics are represented equally
well in coordinate and momentum spaces.  Bound states problems, which
by definition deal with normalizable wavefunctions, can actually be
solved equally well in either space, while scattering problems, which
in the time-independent Schr\"{o}dinger theory deal with
non-normalizable states, are more of a challenge in momentum space.
This challenge arises, in part, because boundary conditions are more
naturally imposed in coordinate space, and, in part, because
non-normalizable states cannot be Fourier transformed\cite{hernandez}.
In spite of the difficulties, momentum-space calculations are
important because fewer approximations are needed there to handle the
nonlocal potentials arising in many-body and field theories.

The Coulomb problem in momentum space has actually been ``solved'' a
number of times---possibly starting with Fock's study of the hydrogen
atom\cite{fock}---yet no one numerical approach appears to provides
the requisite precision for all applications. The real ``problem'' is
that the Coulomb potential between a point projectile (P) and a target
(T),
\begin{equation} \label{vc}
V_{c}({\bf k}', {\bf k}) = \frac{Z_{P}Z_{T}e^{2}}
{2\pi^{2}q^{2}} \rho(q)\ , \ \ \  q = |{\bf k}'-{\bf k}|\ ,
\end{equation}
has a $1/q^{2}$ singularity which must be regularized before a
numerical solution is implemented\cite{qmII}.  (The form factor
$\rho(q)$ in (\ref{vc}) accounts for the finite size of the target's
charge distribution and makes the potential well-behaved at large
$q$---but does not remove the singularity at $q=0$.)  Kwon and
Tabakin\cite{tabakin} solved the bound-state problem with the
potential (\ref{vc}) by using Land\'{e}'s technique\cite{lande} of
subtracting a term from (\ref{vc}), which makes its integral finite,
and then adding in a correction integral.  Alternatively, Ciepl\'{y}
et al.\cite{mach} solved the bound-state problem by using the
Vincent--Phatak (VP) procedure\cite{VP}, which deals with the Fourier
transform of a Coulomb potential which has been cut off beyond some
radius $R_{cut}$,
\begin{equation}
V_{c}^{cut}({\bf k}', {\bf k}) = \frac{Z_{P}Z_{T}e^{2}}{2\pi^{2}q^{2}}
 \left[ \rho(q) - \cos(qR_{cut})\right]\, . \label{vcut}
\end{equation}
While $V_{c}^{cut}({\bf k}', {\bf k})$ is clearly finite as
$q\rightarrow0$, it produces  wavefunctions which must have their
asymptotic behavior corrected.

The VP procedure was originally formulated for intermediate-energy
pion scattering from light nuclei\cite{VP} where it provided
sufficient accuracy\cite{lpott}.  However, the accuracy has become a
concern for intermediate-energy proton scattering where the proton's
much larger mass leads to correspondingly larger momentum transfers
and correspondingly greater numbers of partial waves.  Crespo and
Tostevin\cite{surrey} and Picklesimer et al.\cite{inaccurate2} have
documented difficulties with the VP procedure, difficulties which
appear as a sensitivity of the computed phase shifts to the cutoff
radius, or a several-percent error in the phase shift when compared to
coordinate-space calculations. Both references suggest algorithms to
reduce the errors.  Alternatively, Elster et al.\cite{thaler} applied
the two-potential formula to the Coulomb and nuclear potentials and
outlined an approach requiring multiple, numeric Fourier transforms
between coordinate and momentum spaces. In contrast, Arrellano et
al.'s study of intermediate-energy proton scattering from spinless
nuclei\cite{accurate} simply made the VP procedure sufficiently
precise by using some high-precision partial wave expansions developed
by Eisenstein and Tabakin\cite{eisenstein}. (As a check, they
transformed the potentials to coordinate space and solved the
equivalent integro-differential equation.)

In the present paper we generalize the Blatt--Biedenharn phase shift
parameterization and the VP procedure in order to handle
intermediate-energy proton scattering from spin $1/2$ nuclei in which
tensor forces couple states of differing angular momenta. In
Sec.~\ref{spin0} we reformulate the VP procedure for uncoupled
channels, in Sec.~\ref{spin1/2} we present our new formulation for
coupled channels, and in Sec.~\ref{details} we give some details of
an application to 500 MeV proton scattering from $\mbox{}^{3}$He.

\section{Uncoupled States (0 $\times$ 0, 0 $\times$ 1/2)} \label{spin0}

Vincent and Phatak formulated their procedure in terms of phase
shifts.  We reformulation it in terms of T-matrix elements to avoid
unnecessary conversions between phase shifts and amplitudes.  The
summed nuclear plus cut-off Coulomb potential $V$, can be used in the
Lippmann-Schwinger equation:
\begin{equation}
        T_{l\pm}(k',k) = V_{l\pm}(k',k) + \frac{2} {\pi}
	\int_{0}^{\infty}p^2\,dp
	\frac{V_{l\pm}(k',p)T_{l\pm}(p,k)} {E + i\epsilon -
	E(p)}\ , \label{LS1}
\end{equation}
which is then solved in the normal way since $V_{l\pm}(k',k)$ is
well--behaved in momentum space (short-ranged in coordinate space).
If the ${\bf r}$-space nuclear potential vanishes beyond a range $R$,
and if the Coulomb cutoff-radius $R_{cut}$ is chosen larger than $R$,
then the wavefunction in the intermediate region, $R \leq r \leq
R_{cut}$, can be expressed in terms of a free wave shifted by an
intermediate phase shift $\delta_{l}$\cite{VP}:
\begin{eqnarray}
u_{j=l\pm 1/2}(R \leq r \leq R_{cut})
	& = & N e^{i\delta_{l\pm}} \left[\sin \delta_{l\pm}
\,G_{l}(kr) + \cos\delta_{l\pm}  \,F_{l}(kr) \right]\ , \label{inner}\\
	&\equiv &  N \left[ F_{l}(kr) + \hat{T}_{l\pm} \,H_{l}^{(+)}(kr)
 \right],  \label{FHform}\\
\hat T_{l\pm} &=& e^{i\delta_{l\pm}}\sin\delta_{l\pm}
	=    - \rho_{E} T_{l\pm}\ , \ \ \
	\rho_{E} = 2k_{0}\frac{E_{P}(k_{0})E_{T}(k_{0})}
	{ E_{P}(k_{0})+E_{T}(k_{0}) }\ ,
\end{eqnarray}
where $\delta_{j=l\pm}$ is the intermediate phase shift arising from
the short-range potentials and $N$ is a normalization constant.  Here
$l$ is the orbital angular momentum, $j=l\pm 1/2$ is the total angular
momentum, and we have used  two equivalents forms for the
wavefunctions\cite{qmII}.

The wavefunction for $r>R_{cut}$ can be expressed in terms of a phase
shift $\delta_{l}^{c}$ (the amount a point-Coulomb wave is shifted),
which, in turn, is determined by matching the intermediate
wavefunction to an outer one at $r=R_{cut}$. The outer
wavefunction for $r \geq R_{cut}$ has the same form as the intermediate
one, but with the free waves replaced by Coulomb waves and the
intermediate phase shift $\delta_{l\pm}$ replaced by the phase shift
relative to point-Coulomb waves $\delta_{l\pm}^{c}$:
\begin{equation}
	u_{l\pm}(r \geq R_{cut}) = N'\left[F_{l}(\eta,kr) + \hat T_{l\pm}^{c}
	\,H_{l}^{(+)}(\eta,kr)\right] \ , \ \ \
	\hat T_{l\pm}^{c} =
	e^{i\delta_{l\pm}^{c}}\sin\delta_{l\pm}^{c}\ .
\end{equation}
Here $\hat T_{l\pm}^{c}$ is the T matrix for scattering from short
range forces in the presence of a point-Coulomb force, $\eta =
Z_{P}Z_{T} e^{2}/ v$ is the Sommerfeld parameter, $F_{l}(\eta,kr)$ is
the regular Coulomb function, and $H_{l}^{(+)}(\eta,kr)$ is an the
outgoing wavefunction; for example,
\begin{equation}
H^{\pm}_{l}(\eta,kr) \equiv G_{l}(\eta,kr) \pm i F_{l}(\eta,kr)
\sim exp\{\pm i[kr -l\pi/2 + \sigma_{l} - \eta \ln(2kr)]\}\ .
\label{coulwf}
\end{equation}
The requirement that the logarithmic derivative of $u_{l\pm}(r\leq
R_{cut})$ match $u_{l\pm}(r \geq R_{cut})$ at $r=R_{cut}$ determines
$\hat T_{l\pm}^{c}$ as:
\begin{equation}
\hat T_{l\pm}^{c} = \frac{\hat T_{l\pm}[F_{l}(\eta),H_{l}^{(+)}]
+ [F_{l}(\eta),F_{l}]}{[F_{l},H^{(+)}_{l}(\eta)]
+ \hat T_{l\pm}[H^{(+)}_{l},H^{(+)}_{l}(\eta)]}\ , \label{spinless}
\end{equation}
where the brackets indicate Wronskians.

Finally, the non--spin--flip scattering amplitude can be expressed in terms
of the outer, Coulomb--modified phase shifts
$\delta_{l}^{c}$\cite{lpotp}:
\begin{eqnarray}
f(\theta) &=& f^{c}_{pt}(\theta) + f^{nc}(\theta)\ ,\\
f_{pt}^{c}(\theta) &=& - \frac{\eta} {2 k \sin^{2}(\theta/2)}
exp\{2i\left[\sigma_{0} - \eta\ln\sin(\theta/2)\right]\}\ , \label{fpt}\\
f^{nc}(\theta) &=& \frac{1} {2ik} \sum_{l=0}^{\infty} (2l+1) e^{2i\sigma_{l}}
\left(e^{2i\delta_{l}^{c}} - 1 \right) P_{l}(\cos\theta)\ .
\end{eqnarray}
Here $f^{c}_{pt}$ is the amplitude for scattering from a point-Coulomb
potential and $f^{nc}$ is the amplitude for scattering from the
short-ranged potentials in the presence of the point Coulomb
force.

\section{Coupled States ($\frac{1}{2} \times \frac{1}{2}$)} \label{spin1/2}

If the interaction couples angular-momentum states, we must generalize
the VP method---even if the Coulomb force does not directly couple the
states.  When two non-identical spin $1/2$ particles interact through
a tensor force, the total angular momentum $j$ remains a good quantum
number yet there is coupling within the triplet spin state as well as
between the triplet and singlet states. Accordingly, (\ref{LS1}) is
generalized to the coupled integral equations:
\begin{equation}
        T_{l'l}^{j(s's)}(k',k) = V_{l'l}^{j(s's)}(k',k) + \frac{2} {\pi}
	\sum_{L\,S}\int_{0}^{\infty}p^2\,dp
	\frac{V_{lL}^{j(s'S)}(k',p)T_{Ll'}^{j(Ss)}(p,k)} {E + i \epsilon -
	E(p)}\ , \label{LS2}
\end{equation}
where the sum is over the coupled states.  If the two particles
interact through an optical potential, the phase shifts are complex
and the S matrix non-symmetric. In the Appendix we generalize the
conventional Blatt-Biedenharn NN parameterization
\cite{stapp,blatt,eisenberg} and show that
the S matrix elements have the form:
\begin{eqnarray}
	S_{l'=j\pm1,l=j\pm1}(k_{0}) &\equiv& S_{\pm\pm}
	= \delta_{ll'} - 2i\rho_{E}T_{ll'}(k_{0},k_{0}) \label{t2s}\\
	S_{++} &=& \label{s1}
	(\cos\epsilon_{+-}\cos\epsilon_{-+}e^{2i\delta_{++}}
	+ \sin\epsilon_{+-}\sin\epsilon_{-+}e^{2i\delta_{--}})/\det U \ , \\
	S_{+-} &=&
	(\sin\epsilon_{-+}\cos\epsilon_{-+}e^{2i\delta_{++}}
	- \cos\epsilon_{-+}\sin\epsilon_{-+} e^{2i\delta_{--}})/\det U\ ,
	\label{s2}\\
	S_{-+} &=&
	(\sin\epsilon_{+-}\cos\epsilon_{+-}e^{2i\delta_{++}}
	- \cos\epsilon_{+-}\sin\epsilon_{+-} e^{2i\delta_{--}})/\det U \ ,
	\label{s3}\\
	S_{--} &=&
	(\cos\epsilon_{+-}\cos\epsilon_{-+}e^{2i\delta_{--}}
	+ \sin\epsilon_{+-}\sin\epsilon_{-+}e^{2i\delta_{++}})/\det U \ ,
	\label{s4}\\
	\det U &=& \cos\epsilon_{+-}\cos\epsilon_{-+}+\sin\epsilon_{+-}
	\sin\epsilon_{-+} \ . \label{det}
\end{eqnarray}

To apply the VP procedure to channels coupled by an optical potential,
we 1) transform to a new basis in which there is no channel coupling,
2) match the interior wavefunction to point-Coulomb ones in this basis
for which the S-matrix is diagonal, and then 3) return to the original
basis to calculate the scattering observables. One implementation of
these steps would be to take our S matrix elements computed via
(\ref{LS2}) and (\ref{t2s}), assume they have the forms
(\ref{s1})--(\ref{s4}) in terms of phase shifts and coupling
parameters, and then search for the $(\delta_{--},
\delta_{++}, \epsilon_{+-}, \epsilon_{-+})$ which satisfy
these transcendental equations. The $\delta$'s would be the phase
shifts in the basis in which S is diagonal---even though we never
explicitly transform to that basis.

The implementation we have used is more direct and self-testing.
We compute the solution of (\ref{LS2}),  form the nondiagonal S matrix,
\begin{equation}
	[S] = \left[ \begin{array}{cc} S_{++} & S_{+-}\\
	S_{-+} & S_{--} \end{array} \right] \ ,
\end{equation}
and then explicitly diagonalize it with the similarity transformation:
\begin{eqnarray}
	[S']&=& [U] [S] [U]^{-1} \label{todelt}
	 = \pmatrix{ e^{2i\delta_{++}'} & 0  \cr
	0 & e^{2i\delta_{--}'} \cr} \  , \\*[1ex]
	{[U]} &=& \pmatrix{1 & \frac{S_{+-}} {\lambda_{-} - S_{--}} \cr
	\frac{S_{-+}} {\lambda_{+} - S_{++}} & 1 \cr} ,\ \ \
	 {[U]}^{-1} = \pmatrix{1 & \frac{-S_{+-}} {\lambda_{-} - S_{--}}\cr
	\frac{-S_{-+}} {\lambda_{+} - S_{++}} & 1\cr} \frac{1}{\det U}\ ,
	\\*[1ex]
	\lambda_{\pm} &=& \frac{1}{2} \left[
	S_{++} + S_{--} \pm \sqrt{ \left( S_{++} -
 	S_{--} \right)^{2} + 4 S_{+-} S_{-+}} \, \right]\ .
\end{eqnarray}
We now effectively deduce the intermediate $\delta$ phase shift from
the diagonal elements, do the VP matching with the corresponding
$T_{ll}$'s as in (\ref{spinless}) (this effectively determines the
final phase shift $\delta^{c}$), and then we use the original U
matrix to transform back to the basis in which we calculate
observables:
\begin{equation}
	[S^{nc}]= [U]^{-1}[S'][U] \ .
\end{equation}
Even though the method is guaranteed to diagonalize the S matrix, as
an internal test we check that $\left|S_{ll'}^{nc}\right| \leq 1$ and
that $\left|S_{ll'}^{'}\right| \leq 1$.

\section{Computational Details} \label{details}

We have modified the LPOTp code\cite{lpotp} to include the Coulomb
potential in the different spin channels.  As a first test of our
precision we solved the Lippmann-Schwinger equation with a point
Coulomb potential and checked that our answers reproduced the point
Coulomb phase shifts $\sigma_{l}$ [after removal of the
$\eta\ln(2kR_{cut})$ term in (\ref{coulwf})].  We concluded from this
severe test that 48--64 grid points are required to solve the
Lippmann-Schwinger equation and obtain four--five place precision in
$\sigma_{l}$ [there is enough cancellation that three--place precision
does not reproduce the point Coulomb scattering amplitude $f_{c}^{pt}$
(\ref{fpt})].  After an overall $e^{2i\sigma_{0}}$ is factored out
from the sum in (\ref{fpt}), good agreement for $f_{c}^{pt}$ was found
(indistinguishable from the analytic amplitude on a five-decade
semi-logarithmic plot).

As the next test we computed pure-Coulomb scattering of $415$ MeV
protons from the charge distribution of $\mbox{}^{3}$He. We were able
to obtain essentially perfect reproduction of the Born-approximation
amplitude,
\begin{equation}
	f_{c}^{finite} \simeq f_{c}^{ot}(\theta)\rho(q)\, ,
\end{equation}
which proves that we can include short-ranged effects---in addition to
the long-range Coulomb force---with precision of at least $O(\alpha^{2})$.
To actually obtain this agreement we used 48 grid points in the
solution of the Lippmann--Schwinger equation (\ref{LS2}), and
increased the precision of our partial-wave projection:
\begin{equation} \label{Vproject}
	V_{l} (k',k) = \pi^{2} \int_{-1}^{1}
	V({\bf k}',{\bf k})\,
	P_{l}(\cos\theta_{kk}')\,d(\cos\theta_{kk}')\ ,
\end{equation}
until the partial wave summation,
\begin{equation}
	V({\bf k}',{\bf k}) \simeq
	\frac{1}{2\pi^{2}}\sum_{l}^{l_{max}}
	(2l+1) V_{l}(k',k) P_{l}(\cos\theta_{k'k}) \ ,\label{Vsum}
\end{equation}
reproduced all oscillations present in $V({\bf k}',{\bf k})$. We show
a reproduction of this type in Figure~\ref{fig.V} where the many
oscillations arising from the $\cos (qR_{cut})$ term in the cut-off
Coulomb potential (\ref{vcut}) is evident.  We obtained six-place
reproduction of $V({\bf k}',{\bf k})$ using $l_{max}=$ 48 partial
waves and 96 integration points in the partial-wave projection
(\ref{Vproject}). Ten-place reproduction demanded $l_{max}=96$. We
expect these number to scale as $kR$, and so larger nuclei or higher
energies will require more partial waves and grid points.  For these
calculations we used analytic nuclear form factors\cite{McCarthy},
though we also were successful for $\mbox{}^{13}$C using numerical
Fourier transforms of Wood-Saxon densities\cite{Mefford}.  However,
noise and instability do appear for form factors which fall off slowly in
$q$.

An important requirement on the VP method is that the matching radius,
which we take equal to $R_{cut}$, be larger than the range of the
nuclear force (in order to be able to express the outer wavefunction
as a linear combination of Coulomb waves).  However, increasing
$R_{cut}$ makes the cut-off Coulomb potential more oscillatory and
more difficult to reproduce.  In fact, it was the sensitivity to
changes in $R_{cut}$ which led Ref.\cite{surrey} to search for an
alternative to the matching method.  We find that using $R_{cut} \le
5$ fm produces unstable results (presumably cutting off the nuclear
potential), but, as seen in Figure~\ref{fig.sig}, we obtain stable
results for $6\,\mbox{fm}\leq R_{cut} \leq 10 \,\mbox{fm}$.

In Figure~\ref{fig.sig+A} we compare the nuclear--plus Coulomb cross
section and polarization (solid curves) to those calculated without
Coulomb (dashed curves). The exact handling of the Coulomb
potential is seen to have a significant, although small, effect in the
semilog plot of $d\sigma/d\Omega$, and a more pronounced effect for
$A_{00n0}$.  Not plotted, because they essentially overlap the exact
results, are ones in which the Coulomb potential is handled in
impulse approximation:
\begin{equation}
	f(\theta) \simeq  f^{c}_{pt}(\theta)\rho(q) + f^{n}(\theta)\ ,
\end{equation}
with $f^{n}(\theta)$ the scattering amplitude for pure nuclear
scattering.

\section{Conclusion}

We have extended the Vincent-Phatak procedure for the exact inclusion
of the Coulomb potential in momentum space to calculations of proton
scattering from spin $1/2$ nuclei in which spin--dependent forces
couple angular-momenta states. As part of that extension we also
generalized the Blatt-Biedenharn phase shift analysis for the scattering
of two spin $1/2$ particles to cases where the S matrix is no longer
symmetric.  Although our formulation and calculational procedure is
for a  more complicated spin case, we confirm the
finding of Arrellano et al.\cite{accurate} that the VP procedure can
be made sufficiently accurate for intermediate-energy proton
scattering if high-precision partial-wave
expansions and large numbers of partial waves are used.

\acknowledgements

It is our pleasure to thank Shashi Phatak and Lanny Ray for helpful
discussions and suggestions.  We gratefully acknowledge support from
the U.S. Department of Energy under Grant DE-FG06-86ER40283.

\appendix
\section*{Generalization of Blatt--Biedenharn Convention}

The conventional phase shift analysis must be extended when the
angular momentum channels are coupled. Blatt and Biedenharn did this
first for the mixing of the $l=j \pm1$ states within the
nucleon-nucleon triplet state\cite{blatt,eisenberg} by assuming that
the mixed states have the asymptotic forms:
\begin{eqnarray}
 	\lim_{r\rightarrow \infty} u_{j,l=j-1}(r) &=&
	A_{+} e^{-i[kr-(j-1)\pi/2]}
	-B_{+}e^{i[kr-(j-1)\pi/2]}\ , \\
	\lim_{r\rightarrow \infty} u_{j,l=j+1}(r) &=&
	A_{-} e^{-i[kr-(j+1)\pi/2]}
	-B_{-}e^{i[kr-(j+1)\pi/2]}\ .
\end{eqnarray}
The S matrix is defined by the relation among the A's and B's:
\begin{equation}
	\left[  \begin{array}{l} B_{+}\\B_{-}\end{array}\right]
	= \left[ \begin{array}{ll}S_{++}&S_{+-}\\S_{-+}&S_{--}
	 \end{array} \right]
	 \left[ \begin{array}{l}A_{+}\\A_{-} \end{array}\right] .
\end{equation}
For NN scattering below pion production threshold, S must
be unitary because flux is conserved, and symmetric because all terms
in the Schr\"{o}dinger equation are real. For that  case,
the most general form for S, a unitary and symmetric $2\times2$ matrix, is
given by a similarity transformation with mixing parameter
$\epsilon$,
\begin{eqnarray}
	{[S]} &=& {[U]}^{-1}[e^{2i\Delta}]{[U]} \ ,\label{S}\\
	{[U]} &=& \mbox{$ \left[\begin{array}{ll}
	\cos\epsilon_{j}&\sin\epsilon_{j}\\
	-\sin\epsilon_{j}& \cos\epsilon_{j}
	\end{array}\right]$}, \ \ \
	{[e^{2i\Delta}]} = \mbox{$ \left[\begin{array}{ll}
	e^{2i\delta_{++}}&0\\0& e^{2i\delta_{--}}
	\end{array}\right]$} \ ,\label{U}
\end{eqnarray}
where $\delta_{++} \equiv \delta_{l=j+1, l'=j+1}$ and $\delta_{--}\equiv
\delta_{l=j-1, l'=j-1}$.

When dealing with an optical potential, the S matrix is no longer
unitary---which means the phases shifts become complex, as well as no
longer symmetric---which means there are now two mixing parameters. We
assume (\ref{S}) to be valid with the more general
transformation matrix:
\begin{eqnarray} \label{Unew}
	[U] &=& \mbox{$ \left[\begin{array}{ll}
	\cos\epsilon_{+-}&\sin\epsilon_{-+}\\
	-\sin\epsilon_{+-}& \cos\epsilon_{-+}
	\end{array}\right]$},\ \ \
 	{[U]}^{-1} = \frac{1}{\det U} \mbox{$ \left[\begin{array}{ll}
	\cos\epsilon_{-+}&-\sin\epsilon_{-+}\\
	\sin\epsilon_{+-}& \cos\epsilon_{+-}
	\end{array}\right]$},\\
	\det U &=& \cos\epsilon_{+-}\cos\epsilon_{-+}+\sin\epsilon_{+-}
	\sin\epsilon_{-+} \ .
\end{eqnarray}
This leads to the S matrix elements given in (\ref{s1})-(\ref{s4})
which reduce to the standard, coupled case\cite{stapp,blatt} if
$\epsilon_{+-}=\epsilon_{-+}$, and to the standard uncoupled case if
$\epsilon_{+-}=\epsilon_{-+}=0$.  Stapp\cite{stapp} also gave a
parameterization of the S matrix in terms of the ``bar'' phase shifts
which are, in some cases, more convenient in the parameterization of
data. These bar phases, however, are not the ones introduced here,
and, in fact, do not provide a diagonal representation of the S
matrix.



\begin{figure}
\caption{The nuclear plus Coulomb potentials in momentum space
for the spin triplet state with $m_s=m_{s'}=1$ as a function of the
cosine of the angle between ${\bf k}$ and ${\bf k'}$. The summation
(\protect{\ref{Vsum}}) of partial-wave potentials essentially overlaps
the input potential.}
\label{fig.V}
\end{figure}

\begin{figure}
\caption{The differential cross section for 500 MeV proton scattering
from $\mbox{}^{3}$He. Calculations performed using a cutoff radius in
the range $6\, \mbox{fm} \leq R_{cut} \leq 10 \,\mbox{fm}$ fall within
the two curves. The experimental data are from H\"{a}usser et
al.\protect{\cite{hausser}}.}
\label{fig.sig}
\end{figure}

\begin{figure}
\caption{The differential cross section and analyzing power
(unpolarized target, projectile polarized in normal direction) for 500
MeV proton scattering from $\mbox{}^{3}$He.  The solid curves gives
the exact results using the VP method and the dashed curves gives the
results if no Coulomb force is included. The experimental data are
from H\"{a}usser et al.\protect{\cite{hausser}}.}
\label{fig.sig+A}
\end{figure}

\end{document}